\documentclass[a4paper,fleqn,usenatbib]{mnras}

\def\4u{4U~2206+54}

\usepackage[T1]{fontenc}
\usepackage{ae,aecompl}

\usepackage{graphicx}	
\usepackage{amsmath}	
\usepackage{amssymb}	

\newcommand{\msun}{\ensuremath{{\cal M}}$_\odot$}

\newcommand{\lsun}{\ensuremath{{\cal L}}$_\odot$}
\newcommand{\mass}{\ensuremath{{\cal M}}}
\newcommand{\lum}{\ensuremath{{\cal L}}}
\newcommand{\be}{\begin{displaymath}} 
\newcommand{\ee}{\end{displaymath}} 
\newcommand{\beq}{\begin{equation}} 
\newcommand{\eeq}{\end{equation}}

\newcommand{\kms}{\,km/s\,}

\title[Traceback motion study of HMXB 4U\,2206+54 with {\it Gaia}]{The origin of the high-mass X-ray binary 4U\,2206$+$54/BD\,$+$53\,2790}
\author[V. Hambaryan et al.]{
  V. Hambaryan,$^{1,2}$\thanks{E-mail: vvh@astro.uni-jena.de}
  K. A. Stoyanov,$^{3}$
  M. Mugrauer,$^{1}$
  R. Neuh\"auser,$^{1}$
  W. Stenglein,$^{1}$
  R. Bischoff,$^{1}$
\newauthor
  K.-U. Michel,$^{1}$
  M. Geymeier,$^{1}$
  A. Kurtenkov,$^{3}$
   and A. Kostov$^{3}$
  \\
$^{1}$Astrophysikalisches Institut und Universit\"ats-Sternwarte Jena, Schillerg\"a\ss{}chen 2-3, 07745 Jena, Germany \\
$^{2}$Byurakan Astrophysical Observatory, Byurakan 0213, Aragatzotn, Armenia \\
$^{3}$Institute of Astronomy and National Astronomical Observatory, Bulgarian
Academy of Sciences, \\ Tsarigradsko Chaussee 72,
BG-1784 Sofia, Bulgaria\\
}

\date{Accepted XXX. Received XXX; in original form ZZZ}

\pubyear{2022}
\hypersetup{draft}

\begin{document}
\label{firstpage}
\pagerange{\pageref{firstpage}--\pageref{lastpage}}
\maketitle

\begin{abstract}
  Based on the {\it Gaia EDR3} astrometric parameters and our new systemic radial
velocity of the high-mass X-ray binary 4U\,2206+54/BD+53\,2790, we studied the trace back
motion of the system and propose that it originated in the subgroup of the
Cepheus OB1 association (Age$\sim$4-10 Myr) with its brightest star BD+53\,2820 (B0V; \lum$\sim$10$^{4.7}$\lsun).
The kinematic age of 4U\,2206+54 is about $2.8 \pm 0.4$ Myr, it is at a
distance of 3.1-3.3\,kpc and has a space velocity of 75-100\kms with respect to this member star (BD+53\,2820\,)
of the Cep~OB1 association.
This runaway velocity indicates that the progenitor of the neutron star hosted
by 4U\,2206+54 lost about 4-9\,\msun~during the supernova explosion and the latter one
received a kick velocity of at least 200-350\kms.
Since the high-mass X-ray binary 4U\,2206+54/BD+53\,2790 was born as a member of
a subgroup of Cep~OB1, 
the initially most massive star in the system terminated its evolution
within $\la 7-9$ Myr, corresponding to an initial mass $\ga
32$~\mass$_{\odot}$.
\end{abstract}




\section{Introduction}

It is generally accepted that most stars are formed in compact
groups in gravitationally bound clusters with space densities $>$1~\msun~pc$^{-3}$ \citep{2003ARA&A..41...57L} 
or in extended gravitationally unbound stellar associations with lower space densities $<$0.1~\msun~pc$^{-3}$
\citep{2020NewAR..9001549W}.
  
Star clusters form within giant molecular clouds and remain
embedded in clouds for $\sim$ 2 -- 5 Myr before the combination of massive stellar winds and Supernovae
drive out the gas.
The stars that are left behind after the gas expulsion relax to the new potential and attempt to
return to virial equilibrium \citep[][]{2006MNRAS.373..752G,2007MNRAS.380.1589B}. 

\cite{2020MNRAS.495..663W} 
argue that the formation of OB associations did not follow this scenario
and show that they are formed in-situ as relatively
large-scale and gravitationally-unbound structures.
The OB-associations may contain multiple
groups/cores of young stars, having characteristic
 population of the massive, early spectral O-B type and also containing
 numerous low-mass stars.
 They exhibit some spatial and kinematic
 concentration of short-lived OB stars, a fact first realized by
 \citet{1947esa..book.....A,1955Obs....75...72A}, which provided the first
 evidence that formation of single, double and multiple stars still ongoing in the Galaxy.
 Their dimensions can range from a few to a few hundred pc \citep[for
 recent review see, e.g.,][]{2020NewAR..9001549W}.

However, there is also
a significant number  \citep[10--30\%, see, e.g.,][]{1979ApJ...232..520S,2019A&A...624A..66R} of
young massive stars which are observed in the
Galactic general field and called  ``Runaway stars", a term first introduced by
\citet{1961BAN....15..265B}. Runaway stars are thought to have formed in the stellar
associations
and have been ejected into the general Galactic field by two proposed mechanisms:
dynamical ejection or binary supernova.
The first mechanism, proposed by \citet{1954CoBAO..15....3A} in a Trapezium
type (non-hierarchical) young multiple, dynamically non-stable systems,
was further developed by \citet{1967BOTT....4...86P}.
In contrary, the binary ejection mechanism was first proposed by
\citet{1961BAN....15..265B} to explain the ejection of runaway O and B stars out of galactic
plane. In this scenario the secondary star of a close binary becomes unbound
when the primary explodes as a supernova (SN).
  
However, depending on separation and component masses prior to the
explosion (i.e. phase of mass transfer before the SN, and the subsequent
inversion of the mass ratio) and the
amount of asymmetry involved (i.e. the magnitude of the kick velocity imparted
to the neutron star during the explosion), the binary will either get unbound
(ejecting a single runaway star and neutron star) or it will remain bound
\citep[see, e.g.,][]{1998A&A...330.1047T}.
In case of the latter, its center of gravity will be accelerated and one could
expect to observe a binary system, either as a member of a stellar association
or runaway close binary nearby to a parental stellar group, comprised by a neutron star
and a normal star as High- or Low-Mass X-ray Binary (HMXB or LMXB, respectively),
if the separation is sufficiently small for accretion to occur. Note that the
magnitude of the kick velocity also depends on the evolutionary status of the
pre-explosion close binary system \citep[dynamical stability of mass
transfer to the secondary, see, e.g.,][]{2020A&A...634A..49H}.

Note, also, on the possibility of the so-called two-step-ejection scenario,
i.e. massive binary ejection from star clusters and a second acceleration of
a massive star during a subsequent supernova explosion
\citep{2010MNRAS.404.1564P,2020ApJ...903...43D}.

In this context, it is very interesting to identify the parent stellar group of HMXBs in the
Galaxy \citep[see, e.g.,][]{2001A&A...370..170A,2021arXiv210812918V}.
Recently, the HMXB candidate 1H11255-567 ($\mu^{1}$ and $\mu^{2}$ Cru,
spectral types B2+B5, together with a possible neutron star)
was traced back to the Lower-Centaurus-Crux group, where it could have originated 
up to $\sim 1.8$ Myr ago in a supernova at $89-112$ pc \citep{2020MNRAS.498..899N}.
However, in this system, the neutron star nature of the unseen companion is still uncertain --
it could be instead  a very low-mass M-type star or brown dwarf.

In this work, we concentrate on the kinematic study of the unique HMXB
4U\,2206+54, which has been suspected to contain a neutron star accreting from
the wind of its companion BD\,+53 2790.
This optical counterpart was identified by \citet{1984ApJ...280..688S}, as a early-type star.
Further analysis of many space and ground based observations showed that the
system hosts a neutron star accreting from the wind of its companion, BD\,+53
2790 \citep[see, e.g.,][]{2009A&A...494.1073R,2010ApJ...709.1249F,2018MNRAS.479.3366T},
which also exhibits a radial velocity modulation \citep[see further for
details further and ][]{2014AN....335.1060S}.     
  
The neutron star in the system is probably a magnetar - a class of
rare, strongly magnetized neutron stars. The strength of the surface
characteristic magnetic field is estimated of the order of $B_{S} \sim 2 \times 10^{13}-10^{14}$~G of
this neutron star with the very slow spin period of -- $P_{spin} \sim (5540-5570)$~s and the rapid spin-down rate of
$\dot{P_{spin}}=5.6 \times 10^{-7} s s^{-1}$ \citep[][]{2009A&A...494.1073R,2010ApJ...709.1249F,2018MNRAS.479.3366T}.
Currently, the 4U\,2206+54 is the only known HMXB system hosting a accreting magnetar with or
without a fallback disk \citep{2013IAUS..290...93A,2014ApJ...796...46O}.
The donor star does not meet the criteria for a classical Be V star, but
rather is a peculiar O9 V star with higher than normal helium abundance
\citep[][]{2006A&A...446.1095B} and a double peaked H$_{\alpha}$ emission line, as typical for the decretion disks
\citep[][]{2020A&A...634A..49H}. With an orbital period of 9.5 days, 4U 2206+54 exhibits one of the shortest
orbital periods among known HMXBs.

\section{The birth place of 4U~2206+54}\label{sec:birth}
In order to identify the possible birth place of 4U\,2206+54 one needs to determine its
possible membership to a stellar group either currently or in the past.
The latter also requires to perform their trace back motion study in the
Galaxy to test the concept: 4U\,2206+54 and a stellar group or some of its
members in the past were ``in the same place at the same time''.

\begin{figure*}
  \includegraphics[width=16.0cm]{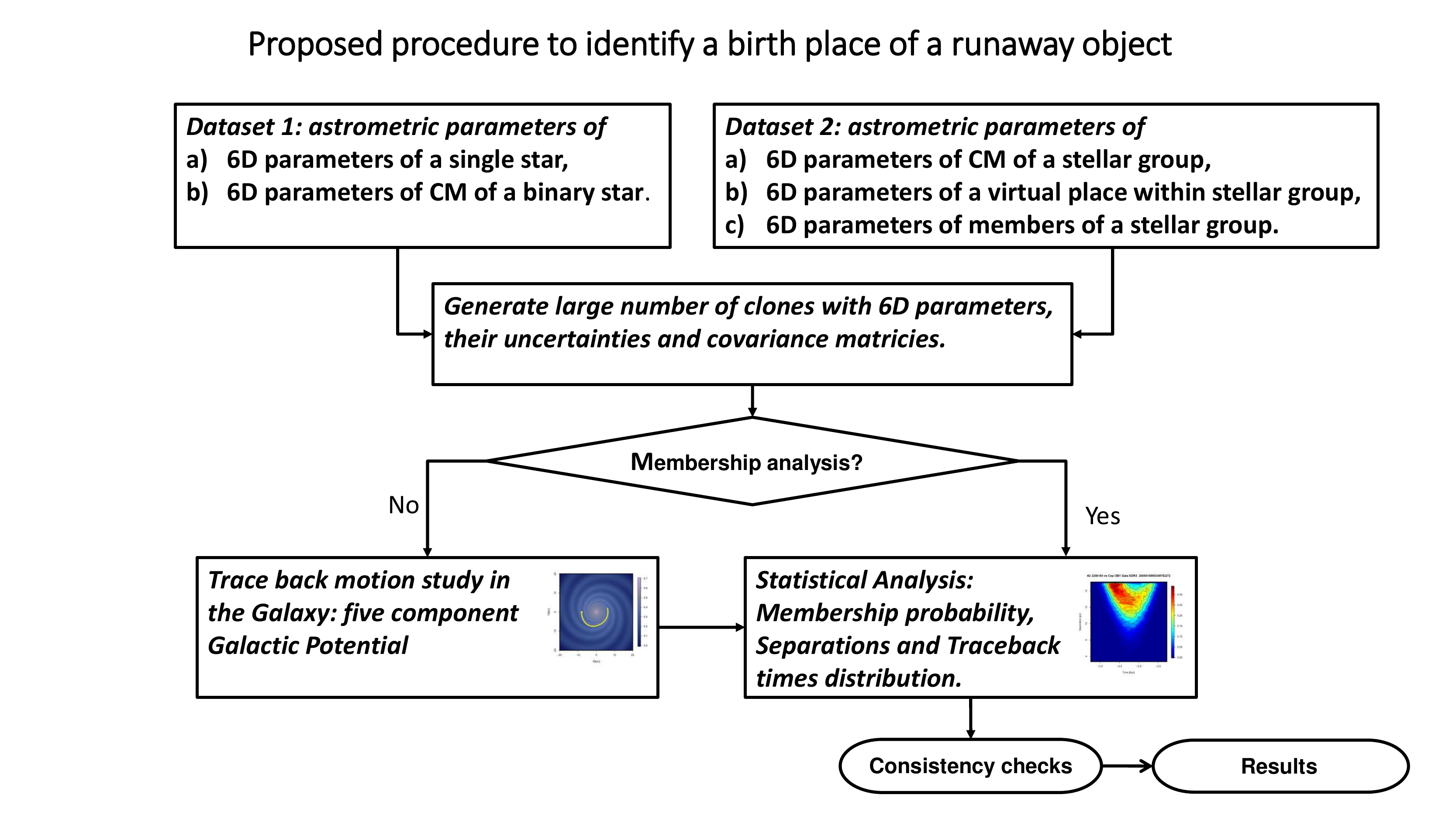}
  \caption{The flowchart of the proposed processing for identification of the
    birth place of runaway object (the concept ``in the same place at the same time'').}
\label{flowchart}
\end{figure*}

It is obvious, that using as an input astrometric and kinematic parameters and their uncertainties of both one
can get, in principle, only certain number of trajectories satisfying some of the
criteria (e.g., minimum separation) of the close stellar passage. In each case, one clearly
gets a probabilistic output \citep[see, e.g.,][]{2000ApJ...544L.133H,2001A&A...365...49H,2010MNRAS.402.2369T,2020MNRAS.498..899N}.
Whether this number is expected from a real pair or by
chance, i.e. occured in the same volume of the space during some time interval in the past,
needs further statistical analysis, given the above mentioned uncertainties of parameters
(see, Fig.~\ref{flowchart}, further Sec.~\ref{subsec:method} and Fig.~\ref{estimtmidmi}).
Finally, further consistency checks must be performed as listed in \citet{2020MNRAS.498..899N},
e.g. that there should not be any more massive (O-type)
star in the host group left that is not yet exploded or that the
flight time should not be larger than the age of a hosting group or neutron star (if known).

First, we have cross-matched the optical companion BD+53\,2790 of the HMXB with 
possible candidate counterparts in \emph{Gaia DR2 and EDR3}  
and identified it with the source 2005653524280214400 \citep[see, also ][]{2021MNRAS.502.5455A}. 

\begin{table*}
\begin{center}
 \caption{The parameters of the optical companion BD $+$53 2790
   of 4U 2206$+$543 and its probable birth counterparts --- the member stars of Cep
   OB1 association (BD $+$53 2820 or HD 235673).}
  \label{spparams}
  \begin{tabular}{lllccccc}
\hline      
    Name     &  \emph{Gaia EDR3}   & Spectral & d$^{**}$ & $\varpi$ & $\mu_{\alpha}cos \delta$ & $\mu_{\delta}$ &  RV$^{***}$  \\
             & Source ID           & type & [pc] &[mas] &  [mas/yr] & [mas/yr] & [km/s] \\ 
\hline
BD $+$53 2790   & 2005653524280214400 & O9.5Vep & 3167.4$^{+165.3}_{-120.1}$& 0.3051$\pm$0.0136  & -4.173$\pm$0.015 & -3.317$\pm$0.014 & -62.7$\pm$8.8 \\
BD $+$53 2820$^{*}$   & 2005418950349782272 & B0IVn & 3545.4$^{+286.8}_{-225.5}$ &  0.2681$\pm$0.0169  & -2.973$\pm$0.018 & -3.350$\pm$0.016  & 15.8$\pm$32.3 \\
HD 235673      & 1981443102866159232 & O6.5V & 4201.6$^{+827.1}_{-489.4}$ & 0.2240$\pm$0.0292 & -3.828$\pm$0.030 & -3.390$\pm$0.026 & -40.0$\pm$10.0  \\ 
\hline
\end{tabular}
\end{center}
\hspace*{-1.6cm}
$^{*}$ Radial velocity of BD $+$53 2820 is variable, may be double--lined
spectroscopic binary \citep{1963ApJ...138.1002A}. \\
\hspace*{-0.8cm}
$^{**}$ Distance estimates are provided by \citet{2021yCat.1352....0B} using parallaxes and
additionally the G magnitudes. \\
\hspace*{-3.1cm}
$^{***}$ Radial velocities and their standard deviations are given according to
the SIMBAD astronomical   \\
database \citep{2000A&AS..143....9W} and corresponding bibliographic entries \citep{1963ApJ...138.1002A,1953GCRV..C......0W}. \\
\end{table*}

Next, we performed a preliminary selection of the possible birth place (i.e. a stellar
group) of HMXB 4U\,2206+54, according to its position and distance, as well as,
upper limits of the age and runaway velocity (e.g., $\sim$~10-20 Myr and $\sim$~100-150~\kms
corresponding to the distance of $\sim$~1-2 kpc), from
the recent catalogues of members of stellar associations
\citep[][]{2017MNRAS.472.3887M,2020MNRAS.493.2339M} and open clusters
\citep[][]{2020A&A...640A...1C}.

\begin{figure*}
\centering
    \vspace{-0.2cm}
    \vbox{
      \hspace{0.2cm}
      \hbox{
        \includegraphics[height=8.0cm]{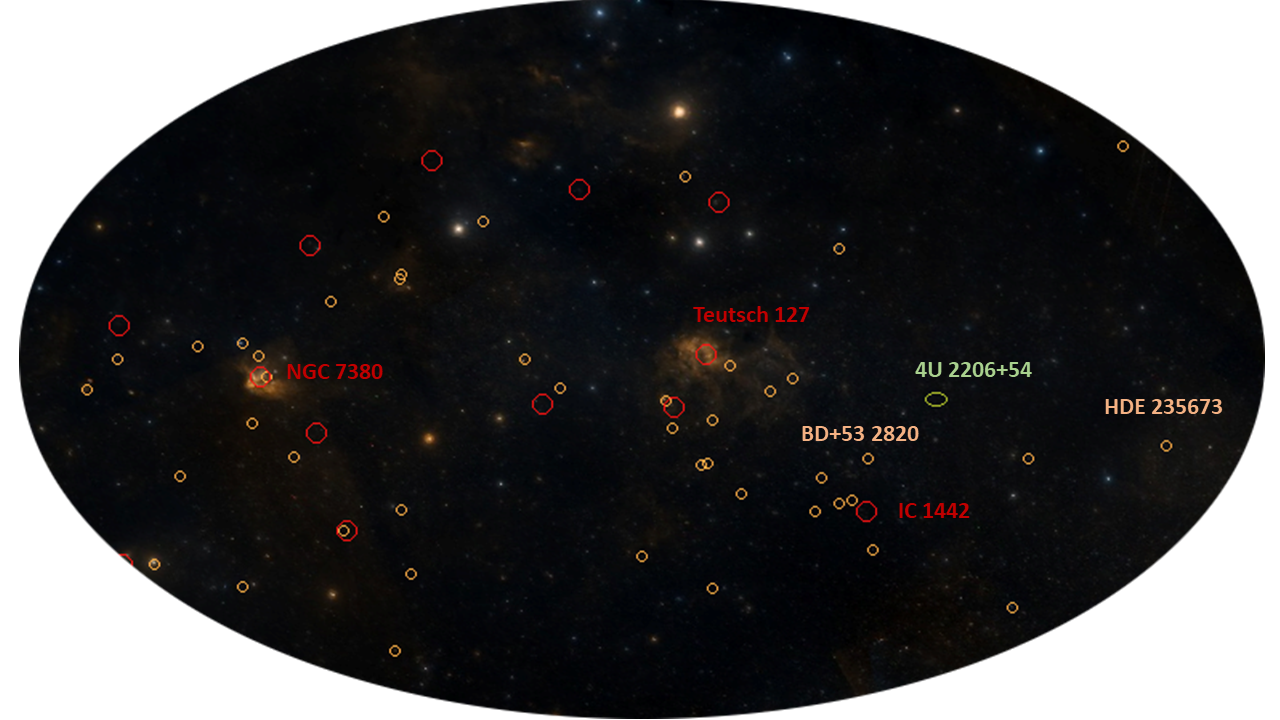}
        }
      \hbox{
        \includegraphics[width=8.6cm]{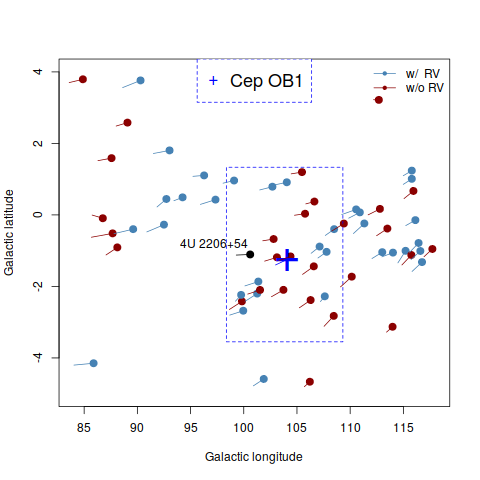}
        \includegraphics[width=8.6cm]{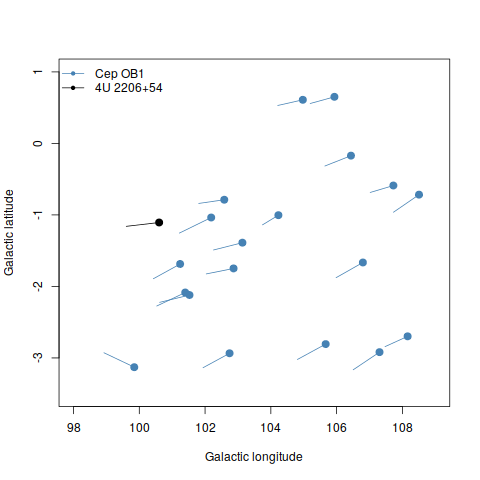}
      }
    }
    \vspace{-0.2cm}
\caption{\emph{Top panel:} Digitized (DSS2) color image of the region of the
  HMXB 4U\,2206+54 (green oval) in the galactic coordinates, prepared with
  Aladin Desktop \citep{2000A&AS..143...33B}.
  The positions of stellar clusters \citep[large red circles,][]{2020A&A...640A...1C} and Cep OB1 association
  member stars \citep[brown circles,][]{2017MNRAS.472.3887M,2020MNRAS.493.2339M} are also indicated. Most relevant objects for this
  study are annotated (for details, see text).
  \emph{Bottom panel:} Galactic positions and proper motions of stellar clusters with Cep OB1 in the center (left
  panel); Cep OB1 association members are shown in the right panel, they can be considered
  as most probable birth counterparts of 4U\,2206+54/BD+53\,2790.}
\label{glgb} 
\end{figure*}

The selection criteria are as follows:
Galatic longitude between 80$^{\circ}$ and 120$^{\circ}$, latitude between -10$^{\circ}$ and 10$^{\circ}$
and distance between 1500~pc and 5000~pc. With this first step of selection the
list consists of 143 stellar clusters and 11 associations.
Taking into account the direction of relative motion of BD+53\,2790 to these stellar
groups (3D or proper motion) and the most probable upper limit of its age
\citep[see, e.g.,][Spectral type~O9.5V, \mass$\sim$ $\ga$15.5~\msun]{2003A&A...404..975M,2012A&A...537A.146E}
the reduced list includes 62 open clusters and only one stellar association (see, Fig.~\ref{glgb})
which can be considered as the probable place of the origin of the HMXB 4U\,2206+54.

For these birth place counterparts, we estimated the membership probability/likelihood of 4U\,2206+54/BD+53\,2790 by comparison
with the bona-fide members of stellar groups given the astrometric and kinematic parameters and
their uncertainties by \emph{Gaia EDR3}. For this purpose we used a
multivariate Gaussian distribution $\mathcal{N}_{n}(\mu,\,\Sigma)$ with
probability density function of Eq.~\ref{eq:probability}) in the five dimensional
space (position, parallax and proper motions)\footnote{Unfortunately, the
  overwhelming majority, in average $\ga $~98\% \citep{2020A&A...640A...1C}, of bona-fide members of the
  stellar groups have no significant number of radial velocity measurements.}:

\begin{align} 
  \begin{split}\label{eq:probability}
  p(z | \mu,\Sigma) = {(2\pi)^{-\frac{1}{2}np}|{\Sigma}|^{-\frac{1}{2}}e^{(z-\mu)\Sigma^{-1}(z-\mu)^T}},
 \end{split}
\end{align}
  
where $z=[\alpha,\delta,\varpi,\mu_{\alpha}cos \delta,\mu_{\delta}]$ is a vector of $np=5$ parameters either of
BD+53\,2790 or bona-fide members of any stellar group with parameters $\mu,\Sigma$. The
corresponding likelihoods $L(Z | \mu,\Sigma) = \prod^n_{i=1}p_{i}$  of BD+53\,2790 or
member stars computed for a large number of generated random vectors with above
mentioned five parameters and their corresponding covariance matrices provided
by \emph{Gaia EDR3}  \citep[][]{2020yCat.1350....0G} and using
$\mathrm{mvtnorm:}$ a Multivariate Normal and t Distributions R
package with Cholesky method, developed by \citet{Genzetal}.   

It turned out, that BD+53\,2790 has a very low
probability to be considered as a member, the logarithm of the
ratio of the mean likelihoods of BD+53\,2790 in
comparison to the members of a stellar group is in the range of -11 to -183.
Note that for the case of Cep OB1 stellar association (the single one in the
list) the logarithm of likelihood ratio is equal to -61.2. 
We obtained similar results (i.e., low and negligible membership
probability) also by using other methods (see, further Sec.~\ref{sec:qnn})
based on the astro-kinematic, as well as photometric parameters of the
bona-fide member stars of stellar groups.  

Hence, we need to study trace back motions of this HMXB and its
above mentioned probable counterparts of the place of origin, i.e.
whether 4U\,2206+54/BD+53\,2790 and a stellar group or one of its member were
in the same place at the same time in the past. 
In order to study the Galactocentric motion of the HMXB 4U\,2206+54 for an input we used
the astrometric parameters of the optical counterpart BD+53\,2790 of the
system presented in \emph{Gaia EDR3}, as well as its systemic radial
velocity. For the latter one, we performed additional spectral observations
and analyzed the combined radial velocity
data set \citep[see further, Sec.\ref{subsec:obs} and][]{1963ApJ...138.1002A,2014AN....335.1060S}.

\subsection{Observational data and analysis (radial velocity)}\label{subsec:obs}

\begin{figure}
\resizebox{\hsize}{!}{\includegraphics{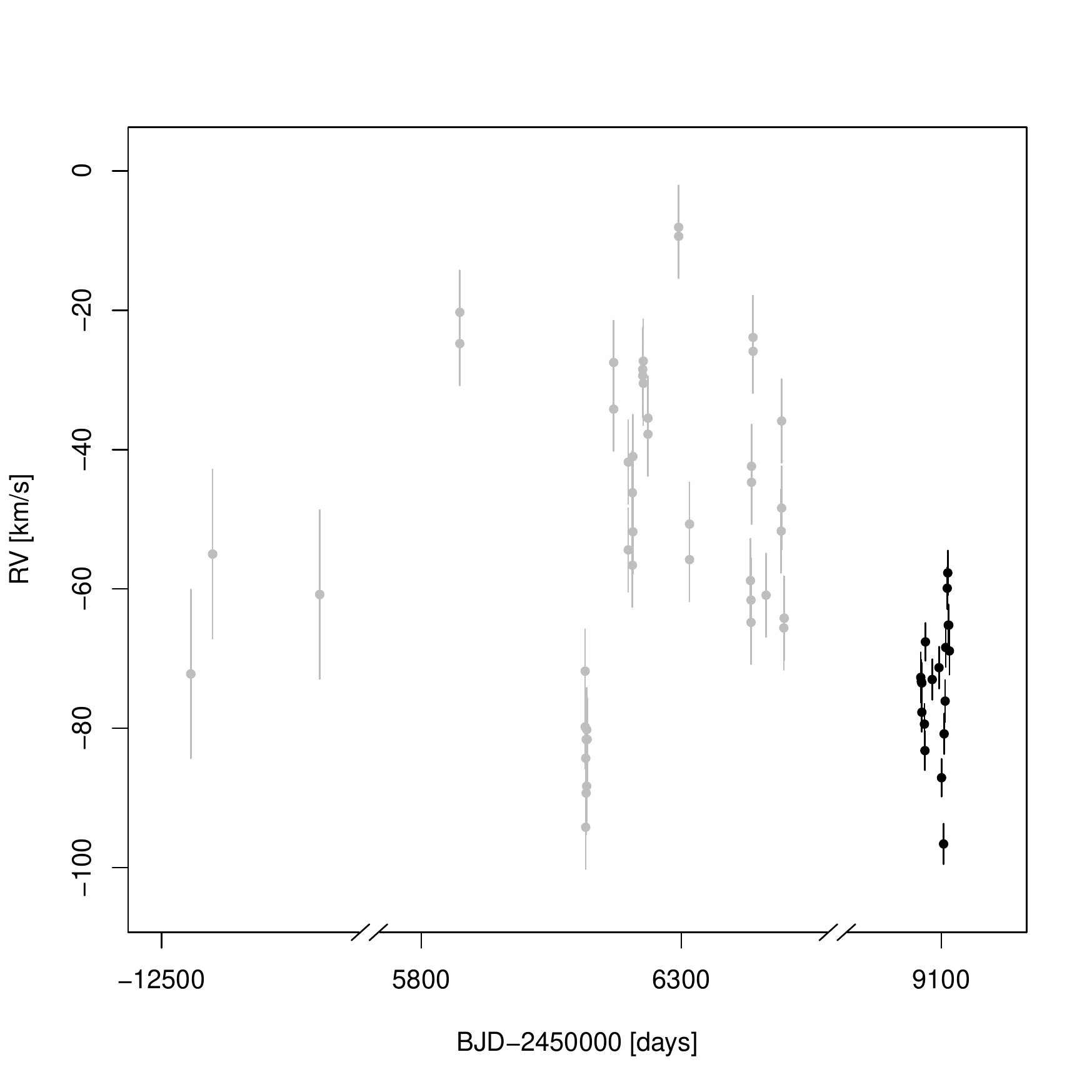}}\caption{The RV
  measurements of BD+53\,2790 from the literature
  \citep{1963ApJ...138.1002A,2014AN....335.1060S} are shown as gray filled
  circles and those derived from our FLECHAS observations with black filled
  circles, respectively. The standard deviations are illustrated as error bars.}
\label{fig_rv}
\end{figure}

We have carried out spectroscopic follow-up observations of the late O9.5Vep
spectral type BD+53\,2790, using the \'Echelle spectrograph FLECHAS at the 90\,cm
telescope of the University Observatory Jena \citep{2014AN....335..417M}. The target was observed in 19
observing epochs between 29 July and 22 September 2020 in the 2x2 binning
mode of the instrument ($<R>=6900$), covering the spectral range between about 3900 and 8100\,\AA.
In each observing epoch three spectra
of the star, each with an exposure time of 1800\,s, were taken always preceded
by three spectra of a ThAr-lamp and of a tungsten-lamp for wavelength- and
flatfield-calibration, respectively. As expected from its spectral type the
spectrum of BD+53\,2790 shows absorption lines of helium and hydrogen.
These spectral lines are broadened and show variations of their profiles
between the individual observing epochs.
The H$\alpha$-line appears in emission and exhibits a prominent central
absorption feature. In addition, several diffuse interstellar bands (DIBs),
as well as the absorption lines of interstellar sodium (Na \textsc{I}
$\lambda$ 5890 \& 5896, alias D$_2$ \& D$_1$) are detected in the spectrum
of BD+53\,2790. The radial velocity (RV) of the target was determined by
measuring the central wavelength of the He \textsc{I} $\lambda$ 5876
(D$_3$)-line, which is the most prominent He-line, present in the
spectrum of BD+53\,2790, which is detected with a sufficiently high
signal-to-noise-ratio ($SNR$), required for accurate RV measurements.
In order to monitor the RV stability of the instrument throughout our
monitoring project, the central wavelengths of the lines of the
interstellar sodium doublet were measured in all spectra, which are
detected in the same spectral order as the D$_3$-line. The RV of the
interstellar sodium lines exhibits a standard deviation of 0.5\,km/s,
consistent with the RV stability of the instrument, reported in other
studies \citep[see, e.g.,][]{2020AN....341..989B} before.
For the RV of BD+53\,2790 we obtain -73.6\,km/s on
average with a standard deviation of 9.4\,km/s (cf. Fig.\,\ref{fig_rv} and Table~\ref{rvstau} with previous RV measurements of the star).
The individual RV measurements are summarized in Table\,\ref{tab_rv}
and are illustrated in Fig.\,\ref{fig_rv}.

\begin{table} \caption{The RVs of BD+53\,2790 for all observing epochs, as determined in our spectroscopic monitoring project together 
with the reached $SNR$, measured in the wavelength range between 5820 and 5850\,\AA.}
\label{tab_rv}
\begin{tabular}{c|c|c}
\hline\hline
BJD-2450000 & RV [km/s]     & SNR\\
\hline
9060.44067 & $-72.7\pm3.6$ & 53\\
9061.45550 & $-73.3\pm3.1$ & 60\\
9062.43062 & $-73.5\pm2.9$ & 71\\
9062.51123 & $-77.7\pm2.8$ & 75\\
9067.48182 & $-79.4\pm2.9$ & 73\\
9068.42210 & $-83.2\pm2.8$ & 62\\
9069.38899 & $-67.6\pm2.7$ & 69\\
9082.47820 & $-73.0\pm2.9$ & 61\\
9095.48532 & $-71.3\pm3.0$ & 67\\
9100.38390 & $-87.1\pm2.7$ & 65\\
9104.34670 & $-96.6\pm2.9$ & 63\\
9105.34096 & $-80.8\pm2.9$ & 87\\
9107.35124 & $-76.1\pm3.0$ & 62\\
9108.34312 & $-68.4\pm2.8$ & 71\\
9111.33612 & $-59.9\pm3.0$ & 65\\
9112.33579 & $-57.7\pm3.2$ & 61\\
9113.32245 & $-65.2\pm3.0$ & 60\\
9114.32943 & $-65.2\pm2.9$ & 64\\
9115.47070 & $-68.9\pm3.4$ & 70\\
\hline
            & $<$RV$>$ $\pm$ SD   & $<$SNR$>$\\
            & $-73.6\pm9.4$       & 66\\
\hline
\end{tabular}
\end{table}

To derive the systemic velocity of the binary system we make use of the
MCMC Bayesian approach and a code developed and provided by
\citet{2005blda.book.....G}, which compares the probabilities of different
models and estimates the parameters of the most probable model.
In the simple model, the difference between the measured radial 
velocities -- $RV_{obs}(t_i)$
and the model predicted ones -- $RV_{model}(t_i)$
at the epoch of $t_i$ \citep[for details, see][]{2007MNRAS.381.1607G}
can be represented as a Gaussian distribution with standard deviation
of $\epsilon(t_i)$: 

\begin{equation}
\label{eq:model}
RV_{obs}-RV_{model}=\epsilon,
\end{equation}

where $\epsilon\, (\epsilon^2 = \sigma^2+s^2)$  includes reported measurement
errors -- $\sigma(t_i)$ and unknown uncertainties -- $s$
(e.g., any real signal in the data) that cannot be explained by the model prediction. 
 
\begin{figure}
\resizebox{\hsize}{!}{\includegraphics{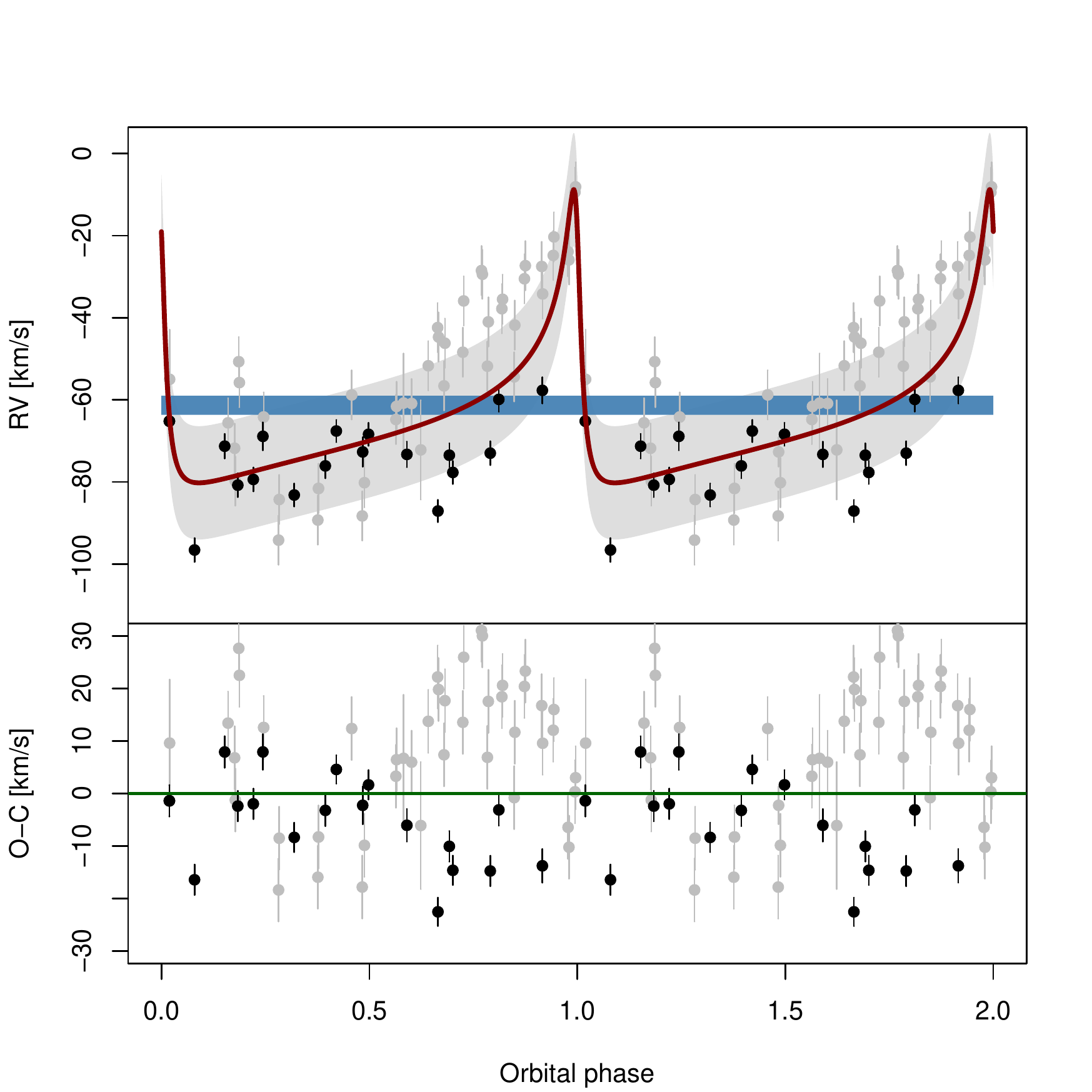}}\caption{The fitted
  binary star model \citep{2005blda.book.....G,2013acna.conf..121G,2017A&A...598A.133D} and residuals of radial velocities
  of BD+53\,2790 from the literature
  \citep[][]{1963ApJ...138.1002A,2014AN....335.1060S} and derived from our
  observations. The red line depicts the most probable radial velocity at
  orbital phase predicted by the fitted binary model. The gray area corresponds to the
  predicted uncertainties. The blue band shows the systemic velocity and its
  uncertainty. Lower panel: Residuals of the observed and model predicted
  radial velocities (for details, see text). The RV
  measurements are shown with filled circles as in Fig.~\ref{fig_rv}.}
\label{fig_rv_fit}
\end{figure}

The best-fitting orbital parameters are listed in Table \ref{Orbit}.
Note that the values of the fitted parameters and their
uncertainties correspond  to the mean values and standard deviations 
of the peak, mean, mode and median of the reported posterior
probability densities \citep[for details, see][]{2005blda.book.....G,2007MNRAS.381.1607G}.
In Fig.~\ref{fig_rv_fit} are plotted the orbital phase folded radial velocity
curve, the best-fitting solution with model uncertainties, and the residuals
of the fit\footnote{The RV plot, presented in \citet{2014AN....335.1060S},
  is inconsistent with the orbital elements (e.g., $\omega$), derived by the
  authors.}.
The systemic velocity $\gamma=-61.5\pm1.55$\kms together with other astrometric
parameters presented in \emph{Gaia EDR3} intended to serve as an input to retrace
its orbits back in time to investigate the probable birth place and
kinematic age of HMXB 4U\,2206+54.
However, statistically significant lack of the good fit (the reduced chi-square $\ga $5,
see, Table~\ref{Orbit}) with a simple keplerian orbit and relatively larger
value of the parameter $s$ indicates on the  
presence either of an unknown signal (e.g. irregular/unstable variation of the
atmospheric layers, mass transfer or rotation, presence of an accretion/decretion disk, etc.)
in the data or small sample size with larger
errors or applied model simplicity. Therefore, to be conservative, for the
study of trace back motion of HMXB 4U\,2206+54 for the input parameter systemic radial
velocity we used a relatively large interval, i.e. the randomly generated $V_{sys}$
values were drawn from Gaussian distribution with the mean value equal to
the fitted systemic velocity $V_{sys} \equiv \gamma=-61.5$\kms with standard
deviation of $SD_{Vsys}=15.0$\kms.

\begin{table}
  \caption{Orbital parameters of 4U\,2206+54}
  \begin{tabular}{l|r}
\hline
Parameter                       & Value\\
\hline
\multicolumn{2}{c}{Fitted Parameters}\\
\hline
$P$ (d)         & 9.55346 $\pm$ 0.001\\
$T_p$ (d)             & 2456227.873 $\pm$ 0.004\\
$e$                     & 0.74 $\pm$ 0.13\\
$\omega$ (deg)          & 48.3 $\pm$ 4.5\\
$\gamma$ (km/s) & -61.50 $\pm$ 1.55\\
$K_1$ (km/s)            & 32.88 $\pm$ 6.29\\
$s$ (km/s)            & 11.83 $\pm$ 2.92\\
\hline
\multicolumn{2}{c}{Derived Parameters}\\
\hline
$a_1\sin i$ ($10^6$ km) & 3.00 $\pm$ 0.36\\
$f(m_1,m_2)$ ($\mass_\odot$)        & 0.0115 $\pm$ 0.004\\
\hline
\multicolumn{2}{c}{Other Quantities}\\
\hline
$\chi^2$                & 315.8\\
$N_{obs}$ (primary)     & 65\\
Time span (d)        & 21558.971\\
$rms$ (km/s)  & 13.60\\
\hline
\label{Orbit}
\end{tabular}
\end{table}

\subsection{Motion of 4U~2206+54 in the Galaxy}\label{subsec:method}

To study the Galactocentric motion of a single point mass (a star, binary or
cluster) we use a numerical integration of its equations of motion in the
gravitational field of the Galaxy expressed in a rectangular Galactocentric
frame. Namely, for the Galactocentric motion of 4U\,2206+54/BD+53\,2790, the possible parental stellar cluster
and association we make use of the code
described in \citet{2020MNRAS.498..899N}, which computes the orbits by a 
numerical integration
of their equations of motion as defined by the Galactic gravitational potential
consisting of a three component (bulge, disk and halo) axisymmetric model
\citep[Model III from][]{2017OAst...26...72B}. In addition, the Galactic
gravitational potential is supplemented with the more realistic,
non-axisymmetric and time dependent terms, which take into account the
influence of the central bar and the spiral density wave 
\citep{1993A&A...274..189P,2008A&A...480..735F,2019MNRAS.488.3474B}.

In order to take account of the uncertainties in the astrometric parameters of
the star and associations, each one was replaced by a large number of clones, each with
astrometric parameters drawn from a multivariate normal distribution.
This is done by making use of the covariance matrix of the astrometric parameters
from \emph{Gaia EDR3} for the star and from a
stellar cluster/association centroid parameters \citep{2020A&A...640A...1C,2018A&A...619A.155S,2017MNRAS.472.3887M,2020MNRAS.493.2339M} or
the astrometric parameters of the individual member star \citep{2018A&A...616A...1G,2020yCat.1350....0G}. 
Such a procedure is superior to the individual,
independent random drawing of each parameter that ignores their 
mutual dependence and result in to the more realistic probability distribution functions of the separation
between 4U\,2206+54 and the centre of stellar group or any member star (see,
e.g., Fig.~\ref{estimtmidmi} and Sec.~\ref{sec:res}).

For numerical integration we utilize the fast and accurate Gauss-Everhart orbit integrator provided
by \citet{Avd}.

Based on the {\it Hipparcos} proper motion of the
HMXB HD153919/4U1700-37 \citet{2001A&A...370..170A} propose that
it originates in the OB association Sco~OB1 within $\la 6$ Myr
(kinematic age being $\tau=2.0~\pm~0.5$ Myr).
Most recently, \citet{2021arXiv210812918V} confirmed that the high-mass X-ray binary
HD153919/4U 1700-37 originates from NGC6231,
the nucleus of the OB association Sco OB1, with its kinematic age of 2.2 Myr,
based on the \emph{Gaia DR2} proper motions and parallaxes.
We applied our approach to this system based on the more precise \emph{Gaia EDR3} data and confirmed
that both the place of origin in Sco~OB1 and the kinematic age of HMXB HD153919/4U1700-37 ($\tau=2.33~\pm~0.05$ Myr).

\section{Results}\label{sec:res}

Our trace back motion study of 4U\,2206+54 and its possible parental
stellar groups (see, Sec.~\ref{sec:birth}) 
revealed that only the association Cep OB1 can be considered as a
candidate. The astrometric and kinematic parameters of its centroid was
determined by member stars \citep[][]{2017MNRAS.472.3887M,2020MNRAS.493.2339M}
present in the \emph{Gaia EDR3} catalogue.
Note that the used distances of member stars and their
uncertainties are provided by \citet{2021yCat.1352....0B} using parallaxes and
additionally the G magnitudes.
 It turned out that the trace back times of the pair
 (i.e. the HMXB 4U\,2206+54 within the association Cep~OB1, $\sim$~150 pc) are
 distributed almost uniformly over a range from 1.3~Myr to 15~Myr.
 Given the fact that Cep~OB1 association has a relatively large size (with a distance $\sim$~2.7-3.5~kpc and
several degrees on the sky), and that it is very elongated in the direction of the Galactic
longitude (see Fig.~\ref{glgb}), suggesting that it may include a chain of OB
associations \citep[][]{2017MNRAS.472.3887M,2020MNRAS.493.2339M} or cores of different ages (see,
Sec.~\ref{sec:qnn}), we performed also a trace back motion study
of 4U\,2206+54 and each member star to identify the most probable common birth place
inside of the Cep~OB1 association. Note that from 58 member stars
\citep[][]{2017MNRAS.472.3887M} of Cep~OB1 46 have an entry in \emph{Gaia EDR3} and only 23 have
also radial velocity measurements.
It turned out that only 2 member stars, HD\,235673 and
BD+53\,2820, with spectral types  of O6.5V and B0V, respectively, show a significant
number of close passages with BD+53\,2790. Namely, from 1 million Monte-Carlo
simulations 1234 (0.12\%) and 52936 (5.3\%) rated as success, i.e. the minimum
separation does not exceed 15~pc within 20 Myr in the past,
accordingly.

\begin{figure*}
    \vspace{-0.2cm}
      \hbox{
        \includegraphics[width=8.6cm]{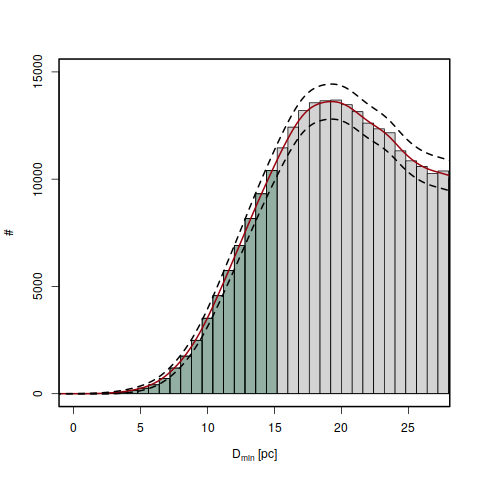}
        \includegraphics[width=8.6cm]{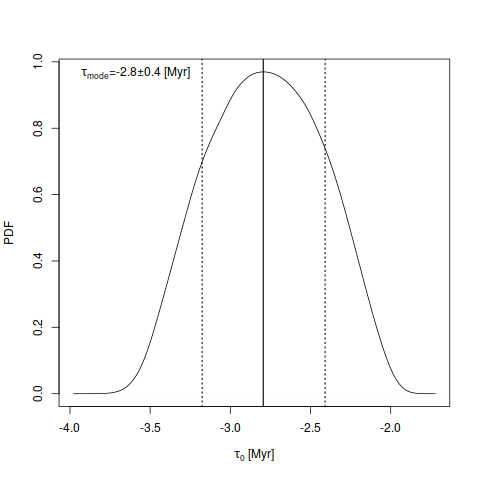}
      }
\caption{Distributions of minimum separations (D$_{min}$) and corresponding
  flight times ($\tau_0$) of closest stellar passage of 4U\,2206+53 and BD+53\,2820 ($\le 15$ pc separation,
  rated as success, marked as filled green area) according to the trace back motion study of
  them in the Galaxy. The red curve with enveloping dashed curves show the fit
  of expected distribution of minimum separations for the 3D case
  \citep[Eq.~A3 in Appendix,][]{2001A&A...365...49H}.  
  The highest posterior density (HPD) interval, 68\% of area, is determined as a probabilistic 
region around a posterior mode of kinematic age of 4U\,2206+53 and depicted as vertical dashed-lines (for
details, see in the text).}
\label{estimtmidmi} 
\end{figure*}

Moreover, the distributions of the trace back times of these ``small'' fractions of
successful cases are unimodal (see, e.g., Fig.\ref{estimtmidmi})
and a significant amount of them, namely 692~($\sim$56\%) and 36929~($\sim$70\%), is concentrated
within relatively narrow time intervals $\delta$t=2.8\,(12.4-15.2)\,Myr and
$\delta$t=0.8\,(2.4-3.2)\,Myr in the past, respectively.

In order to compare the obtained numbers of successful cases with the expected numbers
of cases when our HMXB and a Cep OB1 member
star (4U\,2206+54--BD+53\,2820 or 4U\,2206+54--HD\,235673) in reality were at the same place at
the same time, we created virtual pairs inside Cep OB1 association
at the positions corresponding to BD+53\,2820 and HD\,235673.
We ran them forward with the kinematic properties (proper motions and RVs, see
Table~\ref{spparams}) of flight times from 2.4 to 3.2 Myr and from 12.4 to 15.2 Myr in steps of 0.05 Myr.
For each of the times in the interval, we traced back the pair starting
from their virtual positions and using the kinematic properties (proper
motions and RVs) -- and varying them within their
measurement uncertainties (i.e. according to the covariance matrices provided in
\emph{Gaia EDR3}, including as well corresponding parallax/distance errors) for 1 million trials each. 
For each such trial, we then obtained as usual the minimum distance between pairs.
This procedure thus yields the number of expected close approaches (within
e.g. 15 pc) for the above mentioned time intervals.
As a result, with $95\%$ confidence interval under the assumption of
binomial distribution, we obtained (and, thus, expect at least) close meetings within 15 pc
in 2.3~(2.0-2.7)\% and 0.29~(0.21-0.33)\% cases from of 1 million runs
corresponding to the pairs 4U\,2206+54--BD+53\,2820 and  4U\,2206+54--HD\,235673, accordingly.
Shortly, these fractions can be considered as lower thresholds in favour of the
hypothesis that a pair of HMXB and member star of Cep OB1  were at the same place
during the above mentioned time intervals.

Also, we simulated a large number of random "HMXB''s with mean astrometric and kinematic parameters and their
covariance matrices of neighboring stars of 4U\,2206+54/BD+53\,2790 within 10 arcmin extracted from \emph{Gaia EDR3}
and calculated traced back orbits and compared them with the real trajectories of BD+53\,2820 and
HD\,235673.
It turned out that for such a ``random'' 4U\,2206+54  in one million trials only 8 and 2
cases are successful ones (i.e. separation not exceeding 15 pc) with BD+53\,2820 and HD\,235673 in the trace back
time range of 2.4-3.2 Myr and 12.4-15.2 Myr, respectively, i.e.
with $95\%$ confidence interval under the assumption of
binomial distribution, we expect close meetings within 15 pc
in 0.0008~(0.0003-0.001)\% and 0.0002~(0.00002-0.0007)\% successful cases even
with this conservative randomization.

Thus, statistically the vicinity of both member stars (BD+53\,2820 and HD\,235673) of Cep OB1
association in the past can be considered as probable place of the origin of
the HMXB 4U\,2206+54,
thus indicating the probable coeval formation of the progenitor binary system and one of
these stars. Note that the case of BD+53\,2820 can be considered as more
probable one than the one of HD\,235673 (see, further Sec.~\ref{sec:qnn}).    

\begin{figure*}
    \vspace{-0.2cm}
      \hbox{
        \includegraphics[width=10.0cm]{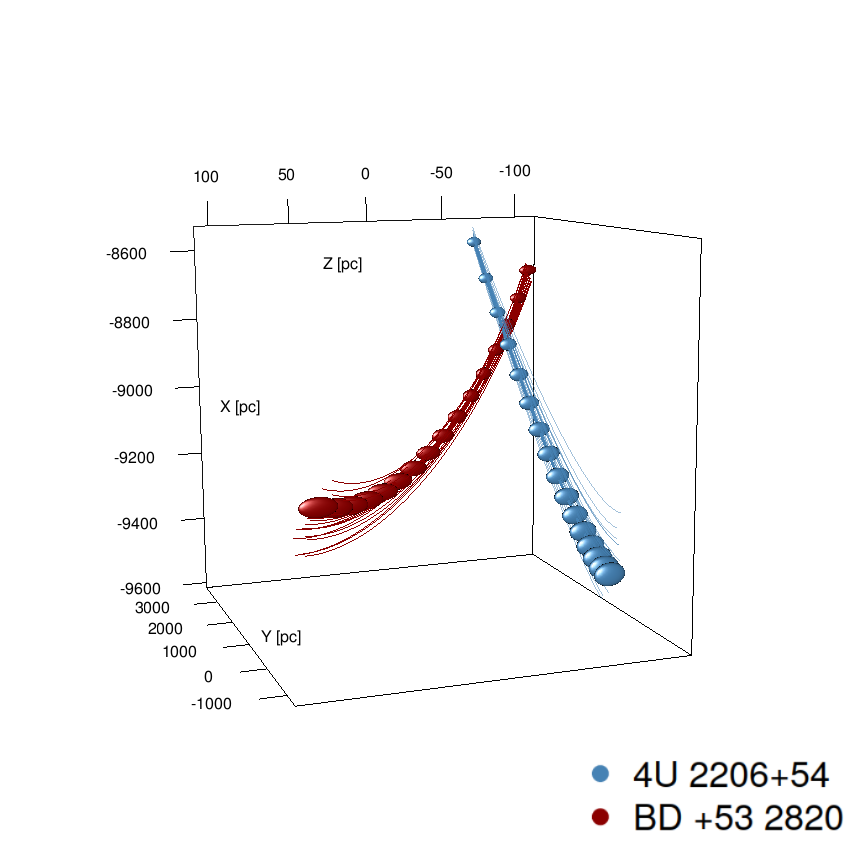}
        \includegraphics[width=8.6cm]{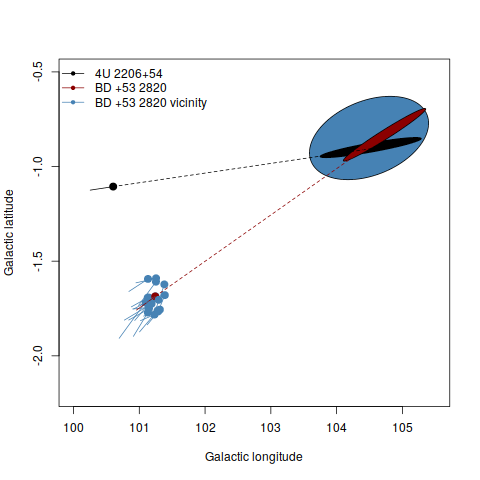}
      }
\caption{\emph{Left panel:} The 3D trajectories of  4U\,2206+54/BD+53\,2790 and BD +53 2820 $\equiv$
  Gaia EDR3 2005418950349782272, a member of Cep OB1 association, in
  Galactocentric Cartesian coordinates in the past. \emph{Right panel:} The
  positions and proper motions of  4U\,2206+54/BD+53\,2790 and subgroup of
  stars in Cep OB1 association with its brightest star BD
  +53 2820 in Galactic coordinates. With filled colors of ellipses are
  indicated the most probable positions of corresponding stars at 2.4-3.2 Myr ago. 
}
\label{fig_trajec} 
\end{figure*}

In Figure~\ref{fig_trajec}, the past 3D trajectories are
displayed for the member star BD+53\,2820 of
Cep~OB1 and for BD+53\,2790\, itself. The analysis of separations and
corresponding times (see, Fig.~\ref{estimtmidmi}) shows that BD+53\,2790 and BD+53\,2820 in
reality  were both inside of the same volume (sphere with radius of $\sim$~15pc) $\tau=2.8~\pm~0.4$ Myr ago.
We observe a similar picture for the neighboring stars of BD+53\,2820 in the
projection on the sky, i.e. purely using position, distance and proper
motions of them (see, Fig.~\ref{fig_trajec}, right panel).

Figure~\ref{estimtmidmi} shows the distribution of the minimum separations,
$D_{\rm  min}(\tau_0)$, and the kinematic ages, $\tau_0$, of the $52\,936$ simulations
mentioned above.

In addition, we studied also the trace back motion of the pair (4U\,2206+54--BD+53\,2820)
with number of input systemic radial velocities corresponding to the observed mean
radial velocity values and standard deviations with different instruments
(see, Table~\ref{rvstau}).
Note that these parameters serving for an input to generate random systemic
velocity are independent of the fitting results and cover a relatively large interval.  

\begin{table}
\begin{center}
 \caption{The input systemic radial velocity parameters (weighted mean and
   standard deviation) for the trace back motion study of 4U 2206$+$54 
   and its success rate, hence the most probable flight time to the subgroup of
   stars (i.e. birth counterpart) of the Cep
   OB1 association with its brightest member star BD $+$53 2820.}
  \label{rvstau}
\begin{tabular}{rrrccr}
\hline      
Number  & RV$_{mean}$ & RV$_{SD}$ & Success  & $\tau_{0}$ & Rem\\ 
of RVs     & (km/s)     & (\kms)      & rate(\%) & (Myr)      & \\ 
  \hline
  3 & -62.7 & 8.8  & 4.4 & $ -2.7^{+0.3}_{-0.5}$ & $^{1}$ \\ 
  43 & -50.6 & 22.5 & 3.8 & $ -2.8^{+0.4}_{-0.1}$ & $^{2}$ \\ 
  19 & -73.6 & 9.4 & 4.2 & $ -2.8^{+0.4}_{-0.4}$ & $^{3}$ \\ 
  65 & -65.7 & 18.9 & 3.9 & $ -2.7^{+0.2}_{-0.5}$ & $^{4}$ \\ 
  65 & -66.6 & 8.7 & 4.3 & $ -2.7^{+0.3}_{-0.4}$ & $^{5}$ \\ 
   \hline
\end{tabular}
\end{center}
$^{1}$ \citet{1963ApJ...138.1002A} \\
$^{2}$ \citet{2014AN....335.1060S} \\
$^{3}$ this work \\
$^{4}$ \citet{1963ApJ...138.1002A,2014AN....335.1060S} and this work \\
$^{5}$ Weighted average of the mean values of the radial velocities measured by
the different RV montoring surveys (instruments).
\end{table}

It turned out that all of these cases confirmed our previous result, i.e. very similar
kinematic age of the 4U 2206 and statistically significant
success rate.

\section{Discussion}\label{sec:qnn}
Based on the parameters of BD+53\,2790 provided by \emph{Gaia EDR3}, we
calculated its absolute magnitude $M_{V}=-4.44 \pm 0.70$ mag \citep[V = 9.84$ \pm $0.2 mag, B = 10.11$\pm$0.19 mag, d = 3135.8$\pm$91.7
pc, Av = 1.8$\pm$0.70 mag,][]{2015A&A...574A..33R} at first. Taking into
account the bolometric correction \citep[BC = -3.2 mag, see, e.g., ][]{2013ApJS..208....9P} for
an O9.5V spectral type star we estimated the mass to be \mass = 23.5$^{+14.5}_{-8.0}$\msun 
using the luminosity-mass relation  for main-sequence stars selected from the
components of detached eclipsing spectroscopic binaries in the solar
neighborhood \citep[][${\rm log}~\lum=(2.726\pm0.203)\times {\rm log}~\mass+(1.237\pm0.228)$]{2018MNRAS.479.5491E}.
Note, the estimate of spectroscopic mass \mass=27$^{+67}_{-23}$\msun \citep[][]{2020A&A...634A..49H}
of BD+53 2790 exhibits 35\% larger than its evolution mass, i.e. the mass of
an object, which exhibits the current stellar and wind parameters, that has
evolved like a single star.
However, in general, there is good agreement between spectroscopic and
evolutionary masses of single stars within the one sigma error
bars \citep[see, e.g.,][]{2014A&A...566A...7N}.

With this initial mass there may be an upper limit for its lifetime in the range of 10-12 Myr 
according to non-rotating and rotating stellar evolution models
\citep{2012A&A...537A.146E,2010A&A...524A..98W,2003A&A...404..975M}.
Hence, the primary of the progenitor of 4U\,2206+54 may have an upper lifetime limit of 7-9 Myr.

Already \citet{1978ApJS...38..309H} lists 11 O-stars within the large Cep OB1
association, which is located at a distance of 3470 pc.
According to \citet{1995ApJ...454..151M} the stellar association Cep OB1/NGC 7380
containing the highest mass stars has formed over a short time span, no
longer than 4-6 Myr.  Despite the fact that most of the massive stars are born during
a period of $\Delta \tau <$3 Myr in this association, some star formation has
clearly preceded this event, as evidenced by the
presence of evolved ($\tau \sim$~10 Myr) 15 \msun~stars
\citep[][]{1995ApJ...454..151M}.

Based on the {\it Gaia} data  \citet{2017MNRAS.472.3887M,2020MNRAS.493.2339M}
studied  the kinematics of OB-associations
with the use of the Tycho-Gaia Astrometric Solution (TGAS) and {\it Gaia DR2} 
and listed 58 member stars of the
Cep OB1 association, having luminosity classes in the range of I to V, with
spectral types of O5-M4, out of which 37 have O-B2 classes, 3 red and 2
evolved A class supergiant stars.  
On the other hand, \citet{2005A&A...438.1163K,2005A&A...440..403K}
identifies 3 ionising star clusters related to the
Cep OB1 association: NGC 7380, IC 1442, and MWSC 3632.
In addition \citet{2017MNRAS.472.3887M,2020MNRAS.493.2339M} included 6
stars of NGC 7235 \citep[9.3~Myr old,][]{2020A&A...640A...1C} with spectral
types of B0-B2 in the list of bona-fide member stars of the
Cep~OB1 association.
Moreover, according to the most recent catalogues of stellar groups \citep[][]{2018A&A...619A.155S,2020A&A...640A...1C}
in the region of Cep OB1 there are more groups in the age range of 4-10 Myr
(see, Fig.~\ref{glgb}).

In order to obtain more constraints on the age of the Cep OB1 association or
its subgroups, we performed a membership analysis of the above mentioned bona-fide
member stars.
First of all, we used {\it Gaia} EDR3 astrometric data and
utilised the UPMASK \citep[Unsupervised Photometric Membership Assignment in Stellar Clusters; ][]{2014A&A...561A..57K}
method to calculate membership probabilities of the observed stars.
The application of this method to 46 \emph{Gaia} EDR3 stars showed that 
a overwhelming majority ($\sim$~85\%) of them a have membership probability $\ge$~0.5,
i.e. they have a common origin in a five-dimensional astrometric space
($\alpha$, $\delta$, $\varpi$/distance, $\mu_{\alpha}\cos\delta$,
$\mu_{\delta}$) in
comparison to the field stars which are spatially
randomly distributed objects of different origins.

We obtained similar results by making use of non-parametric
\citep[e.g., Clusterix 2.0;][]{2020MNRAS.492.5811B} 
and parametric \citep[e.g., BANYAN-Sigma][]{2018ApJ...860...43G} methods, where
also Cartesian 3D (XYZ) positions, kinematic and photometric parameters of
these \emph{Gaia EDR3} stars were used as input parameters.
  
We estimated the lower limit of the age of the Cep OB1 association to be $\sim$~4~Myr from
its turn-off point (HD 235673 $\mathrm{M_V \approx -5.5}$, HD 215835 $\mathrm{M_V
  \approx -6.5}$) in the unreddened absolute visual magnitude-color diagram as a coeval star forming region.

In addition, to updating this age inferred from the above mentioned approach
(i.e. stellar evolution models),
we also tried to assess the kinematic age of the Cep OB1 association as a whole expanding
stellar system. We analyzed how the mean, median or mode of the distribution of mutual distances of
the member stars changes with time.
We performed this by tracing back the orbits of the individual member stars of
the Cep OB1 to determine when they were closest together.
We find that the average distance between the members remains roughly constant for
$\sim$10-15~Myr.
Thus, our empirical approach to estimate the kinematic age of an expanding
(as a whole) stellar system by the analysis of the distribution of mutual distances of
bona-fide member stars in the past \citep[see, e.g.][]{2021MNRAS.500.5552B},
does not show a global minimum,
suggesting also on a possible non-coeval star formation in
this extended Cep OB1 association as a complex star forming region unlike
to a compact ones \citep[see, e.g.,][]{1991SvA....35..135S}.

Thus, the estimated ages of Cep OB1 and 4U\,2206+54 already are excluding HD\,235673 as a
birth counterpart owing to the longer flight time ($\tau=\mathrm{13.2}_{-0.8}^{+2.0}$~Myr, Sec.~\ref{sec:res}).
Moreover, if this O6.5V spectral type star and the progenitor of 4U\,2206+54  
were born together then for the primary mass we would expect at least
40~\msun and maximum lifetime of 4-10~Myr, much shorter than the flight time of
4U\,2206+54 and HD\,235673 to the hypothetical place of the common origin.   

Thus, $\tau=\mathrm{2.8}\pm\mathrm{0.4}$~Myr can be considered
as the most probable kinematic age of 4U\,2206+54,
which suggests a coeval formation of the progenitor binary system of
that HMXB and a subgroup of stars from Cep OB1
association with its brightest member BD+53\,2820.

An application of the UPMASK method to the stars extracted from {\emph Gaia}
EDR3 around the brightest member star BD+53\,2820 in the circle within a radius of
10~arcmin revealed 22 other stars which can be considered as members of this
subgroup. Unfortunately, all of them are too faint and there is no
information about their pre-main sequence nature (e.g. detailed spectral
analysis, X-ray observations) in the astronomical
literature, only two young ytellar candidate objects have distances not
exceeding 500pc from the Sun.
Nevertheless, based on their location the CMD diagram (assuming similar
interstellar absorption $A_{V}\sim 1.2$ as the brightest member BD+53\,2820,
$M_{V}=-3.5 \pm 0.5$, \mass = 17.5$^{+10.5}_{-5.0}$\msun)
showed that the age of the  subgroup can be estimated to be $\sim$~7-10 Myr.
We obtained similar constraint by using  isochrones from the
Geneva stellar models \citep[][]{2012A&A...537A.146E}.

Having estimates of the age range of Cep OB1, the conservative lifetime of the
donor star of the HMXB BD+53\,2790 and the flight time to the probable birth place, we  
estimated the upper limit of the lifetime and hence, the initial mass of the
primary before the SN for all models provided by
\citet{2012A&A...537A.146E,2003A&A...404..975M} to be \mass$_{initial}\sim$~32-60~\msun.

It is difficult to reconstruct the evolution of the massive binary
before the SN explosion. Nevertheless, with our results for the
kinematic age and the orbital parameters of 4U\,2206+54 we may put some constraints on it.

If we consider a circular pre-SN orbit, when the progenitor explodes in a
symmetric SN, an amount of mass is ejected instantaneously.
Ignoring the effects of the impact of the ejected shell on the companion star
and assuming that there is no mass loss or mass transfer during the
circularization of the orbit by the tidal force, the orbital period of
the re-circularized orbit is \citep{1999A&A...352L..87N}: 

\begin{equation}
P_{re-circ} = P_{postSN} (1 - e^2_{postSN})^{3/2},
\label{recirc-P}
\end{equation}
where P$_{postSN}$ and e$_{postSN}$ are the post-SN orbital period and eccentricity respectively.
Using P$_{postSN}$ = 9.56$\pm$0.001~d and e$_{postSN}$ = 0.74 $\pm$ 0.13
we estimated P$_{re-circ}$ = 2.9 $\pm$ 1.8~d.

From the conservation of momentum and the Kepler's Third law,
the runaway velocity $\vartheta$ can be estimated as: 

\begin{equation}
\vartheta = (2\pi G)^{1/3}~\Delta \mass~\mass_1~P^{-1/3}_{re-circ}~(\mass_1 + \mass_2)^{-5/3},
\label{vsys}
\end{equation}
where \mass$_1$ and \mass$_2$ are the masses of the present-day donor star (primary)
and compact object (secondary) respectively. $\Delta$\mass\, denotes the mass of the ejected
material during the SN event which can be estimated using the relative
velocity of 4U\,2206+54/BD+53\,2790 with respect to BD+53\,2820. 

Our analysis of motion shows that 4U\,2206+54
originates in the OB
association Cep~OB1, from which it escaped about 2.8$\pm$0.4~Myr ago due to the
SN of 4U\,2206+54's progenitor. Using parameters of calculated 36\,929 traced
back orbits for the relative
space velocity one obtains $\vartheta \equiv \mathrm{V}_{\mathrm{relative}}$ = 92.6$^{+14.6}_{-16.2}$~\kms with respect
to BD+53\,2790 or its vicinity stars and hence,
$\Delta$\mass~= 5.6$^{+3.6}_{-2.2}$ \msun~for the neutron star of mass
\mass$_2$ = 1.4~\msun. Note that the estimate of $\Delta$\mass\, is not changing
significantly depending on the mass of a neutron star (1.2-2.2~\msun)
and/or period of the re-circularized orbit (i.e. the post-SN eccentricity
$\approx$ 0, at
most by factor of 1.5, cf., Eq.\ref{vsys}). 
Thus, at the moment of the SN instantaneous explosion the collapsing
core would have a mass of 7.0$_{-2.6}^{+4.2}$~\msun, which explodes as a
SN, becomes a neutron star or black hole, and receives a velocity kick,
due to any asymmetry in the explosion. Evidence for such a kick for
non-disrupted systems are large eccentricities of X-ray binary systems \citep[see,
e.g.,][]{1996Natur.381..584K} or observed velocities of radio pulsars
\citep{1994Natur.369..127L}. Clearly, the state of the binary after the SN
depends on the orbital parameters at the moment of explosion and the kick
velocity. For the case of 4U\,2206+54 we estimated the required minimum
kick velocity of a typical neutron
star \citep[Eq. A14 in Appendix,][]{2002MNRAS.329..897H} $\sim$~200--350~\kms 
for the simple case, i.e. imparted in the orbital plane and in the direction of
motion of the pre-SN star, for parameters of the mass range of BD+53\,2790, mass of
the ejected material $\Delta$\mass, orbital velocity (465-530~\kms) of the binary at the moment of
explosion and post--SN runaway systemic velocity ($\mathrm{V}_{\mathrm{relative}}$) of
4U\,2206+54. Note that the above estimated kick velocity of a neutron star is
compatible with kick velocities expected from a unimodal or bimodal Maxwellian
distribution of pulsars \citep[see, e.g.,][]{2005MNRAS.360..974H,2020MNRAS.494.3663I}.

On the other hand, the evolution of massive close binaries is driven by case B mass transfer
\citep{2000A&A...364..563V}. 
In this case, the mass transfer starts after the primary star has finished its
core-hydrogen burning, and before the core-helium ignition.
Resulting from the mass transfer, the remnant of the primary star is its helium core,
while its entire hydrogen-rich envelope has been transferred to the secondary star, which
became the more massive component of the system \citep[conservative mass transfer as
the dominant mode, see, e.g.,][]{2000A&A...364..563V}.
Following \citet{1985ApJS...58..661I}
for the initial mass ($\geq$32~\msun) of a star that will explode as a
SN with helium core mass \mass$_{He}\geq$13.4~\msun and
\mass$_{lost}\geq$6.4~\msun
\citep[the fraction of mass lost $\sim$~0.2][]{2000A&A...364..563V}.

\section{Conclusions}

We presented the following study and results:
\begin{itemize}
\item We found that the member star of Cep OB1 association BD+53 2820 (spectral
  type B0 and luminosity class IV) and runaway
  HMXB 4U\,2206+54/BD+53\,2790 pair satisfies all our criteria for a close meeting
in the past, namely they were at the same time ($2.8
\pm 0.4$ Myr ago) at the same place (distance of $3435 \pm 67$ pc).
It is therefore most likely, that at this location and time, a SN in a
close massive binary took place and can be considered as the place and
time of the origin of the currently observed  HMXB.
For the HMXB 4U\,2206+54/BD+53\,2790, we obtained a runaway velocity of
75-100\kms at the moment of SN explosion.
Our conclusions hold for a wide range of radial velocity of BD+53\,2820  of $23 \pm 16$ km/s.
\item   Given current orbital parameters of the HMXB 4U\,2206+54/BD+53\,2790 and using approaches described by
  \citet{2000A&A...364..563V,1999A&A...352L..87N,1998A&A...330.1047T,2002MNRAS.329..897H,2014LRR....17....3P}
  we estimated a number of parameters of the progenitor binary system, i.e.
mass of the SN progenitor: $\ga $32~\msun (\mass$_{He}\geq$13.4~\msun, \mass$_{lost}\geq$6.4~\msun),
mass of the ejected SN shell $\Delta$\mass $\ga $5 \msun,
required minimum kick velocity of the produced neutron star v$_{kick}\sim$200-350 km/s.
\end{itemize}

\bigskip

\noindent {\bf Acknowledgments.}
We thank the anonymous referee for constructive and useful comments.
We acknowledge financial support from the Deutsche Forschungsgemeinschaft in
grant NE 515/61-1. KAS acknowledge support from grant K$\Pi$-06-H28/2
08.12.2018 "Binary stars with compact object" (Bulgarian National Science
Fund). AK acknowledges support from the Bulgarian Ministry of Education and
Science under the National Research Programme "Young scientists and
postdoctoral students" approved by DCM \#577/17.08.2018.
This work has made use of data from the European Space Agency (ESA) mission
{\it Gaia} (\url{https://www.cosmos.esa.int/gaia}), processed by the {\it Gaia}
Data Processing and Analysis Consortium (DPAC,
\url{https://www.cosmos.esa.int/web/gaia/dpac/consortium}). Funding for the DPAC
has been provided by national institutions, in particular the institutions
participating in the {\it Gaia} Multilateral Agreement.
This research has made use of the SIMBAD database and "Aladin sky atlas"
operated and developed at CDS, Strasbourg
Observatory, France \citep{2000A&AS..143....9W,2000A&AS..143...33B}.
Based on observations obtained with telescopes of the University Observatory Jena, which is operated
by the Astrophysical Institute of the Friedrich-Schiller-University.

\bibliographystyle{mnras}

\section*{Data Availability}
The data underlying this article are available either in the article or from
the Gaia Archive at \url{https://gea.esac.esa.int/archive/}.
Data resulting from this work will be made available upon reasonable request.

\bsp	
\label{lastpage}
\end{document}